\documentclass[12pt,preprint]{aastex}

\begin{document}

\title {The Peculiar Periodic YSO WL 4 in $\rho$ Ophiuchus}

\author{Peter Plavchan\altaffilmark{1,2}, Alan H. Gee\altaffilmark{1}, Karl Stapelfeldt\altaffilmark{2}, Andrew Becker\altaffilmark{3}} 
\altaffiltext{1}{Infrared Processing and Analysis Center, California Institute of Technology, M/C 100-22, 770 S Wilson Ave., Pasadena, CA 91125}
\altaffiltext{2}{Jet Propulsion Laboratory, California Institute of Technology, MS 183-900, 4800 Oak Grove Drive, Pasadena, CA 91109}
\altaffiltext{3}{Astronomy Department, University of Washington, Seattle, WA 98195}

\begin{abstract}
We present the discovery of 130.87 day periodic near-infrared flux variability for the Class II T Tauri star WL 4 (= 2MASS J16271848-2429059, ISO-Oph 128).  Our data are from the 2MASS Calibration Point Source Working Database, and constitute 1580 observations in J, H and K$_s$ of a field in $\rho$ Ophiuchus used to calibrate the 2MASS All-Sky Survey.  We identify a light curve for WL 4 with eclipse amplitudes of $\sim$0.4 mag lasting more than one-quarter the period, and color variations in J-H and H-K$_s$ of $\sim$0.1 mag.    The long period cannot be explained by stellar rotation. We propose that WL 4 is a triple YSO system, with an inner binary orbital period of 130.87 days.  We posulate that we are observing each component of the inner binary alternately being eclipsed by a circum-binary disk with respect to our line of sight.  This system will be useful in investigating terrestrial zone YSO disk properties and dynamics at $\sim$1 Myr.   \end{abstract}


\section{INTRODUCTION}

When stars form and contract onto the main sequence, remnant material can remain in a circumstellar disk.  More than half of solar-type young stellar objects (YSOs) possess these primordial disks from which planets can eventually form \citep[]{meyer97}.  The physical mechanisms responsible for the evolution and dissipation of primordial disks aren't directly observed.  $\rho$ Ophiuchus ($\rho$ Oph) is a $\sim$135 pc star-forming region containing several hundred $\sim$1 Myr YSOs \citep[][]{mamajek08,natta06,barsony05,lada87,lada84}.  Photometric variability is a common property of YSOs, and several large-sky and targeted variability studies of YSOs have been undertaken in the near-infrared \citep[near-IR,][]{oliviera08,barsony97,carpenter01,carpenter02}.  With the Two Micron All-Sky Survey (2MASS) Calibration Point Source Working Database \citep[Cal-PSWDB,][]{skrutskie06}, we are carrying out a program to study the near-IR variability of YSOs in $\rho$ Oph as a probe of stellar and circumstellar disk evolution.  In this letter, we present the discovery of a YSO with long-term periodic variability that we attribute to eclipses by a circum-binary disk.

WL 4 is a previously unremarkable $\sim$1 Myr Class II T Tauri star in $\rho$ Oph \citep[]{natta06,strom95}.  For WL 4, \citet[]{natta06} estimate a J-band extinction of A$_J$=5.5, an effective stellar temperature of 3715 K, an intrinsic luminosity of 1.2 L$_\odot$, and a mass of 0.45 M$_\odot$. \citet[]{natta06} derive only upper limits for accretion from the Pa$\beta$ and Br$\gamma$ emission line equivalent widths, implying WL 4 is not a strong accretor ($\log(\dot M_{acc}/(M_\odot/yr))$$<$$-9.1$).  For comparison, \citet[]{strom95} derived an extinction A$_J$=5.0 and $\sim$1 Myr mass of 1.4 M$_\odot$ from \citet[]{dantona94} isochrones. \citet[]{tsuboi00} identify three consecutive X-ray flares with a quasi-period of $\sim$20 hours.  \citet[]{ratzka05} report a 0.$^\prime$$^\prime$176 companion from high angular resolution imaging, with a projected separation of $\sim$24 AU and a flux density ratio of 0.602$\pm$0.062.  Observations of WL 4 at infrared and sub-mm wavelengths are presented in ${\S}$2.2.

\section{OBSERVATIONS}

\subsection{2MASS Calibration Observations}

2MASS imaged the entire sky in three near infrared bands between 1997 and 2001.  Photometric calibration for 2MASS was accomplished using hourly observations of 35 selected calibration fields, with different fields visited each hour.  One of these fields in $\rho$ Oph covers a region 8.5$^\prime$ wide in R.A. by 60$^\prime$ long in dec., centered at (R.A.,Dec)=(246.80780$^\circ$,-24.68901$^\circ$).  The calibration fields were observed using the same ``freeze-frame'' scanning strategy used for the main survey that yielded a net 7.8 sec exposure on the sky per scan, with six scans of the field taken in alternating declination directions during each hourly calibration observation.  In 3 $\sim$6 month visibility windows spanning 901 days, 1582 independent scans were made of the field in $\rho$ Oph, including 1580 detections of WL 4.  The raw imaging data from each scan of a calibration field were reduced using the same pipeline used to process the main survey.  All source extractions from all scans were loaded into the Cal-PSWDB  \citep[][]{cutri06}.  We do not identify the cause of the two missing detections for WL 4 in the Cal-PSWDB, but it is likely an artifact of the automated processing.

\subsection{Thermal Infrared and Sub-Millimeter Photometry}

The Cores to Disks (c2d)  \textit{Spitzer} Space Telescope Legacy program surveyed star-forming regions including $\rho$ Oph \citep[]{evans03}.  The final c2d data delivery (DR4) includes measured flux densities from observations of WL 4 with MIPS with two epochs at 24 $\mu$m, one epoch each at 70 $\mu$m and 160 $\mu$m \citep[][]{padgett08,rieke04}, and two epochs with IRAC at 3.6, 4.5, 5.8 and 8.0 $\mu$m \citep[][]{allen08,fazio04}.  \citet[]{padgett08} notes that WL 4 and ISO-Oph 129 fall within an elliptical ring of 24 $\mu$m emission $\sim$1 arc-minute in diameter centered on ISO-Oph 125 and 124.  \citet[]{barsony05} observed WL 4 at 10.8 $\mu$m and 12.5 $\mu$m.  The detection at 12.5 $\mu$m appears to be inconsistent with the other measurements.    Observations at 70 $\mu$m with MIPS, and in the sub-mm as reported in \citet[]{andrews07,stanke06}, are confusion limited by emission from ISO-Oph 124,125, and 129.   In our analysis, we treat these photometry as upper limits.


\section{ANALYSIS AND RESULTS}

\subsection{Periodic Variability}

For WL 4, periodic variability is apparent in the unphased data, alternating between bright and faint states.  We identify a period of 130.87$\pm$0.40 days using both the Lomb-Scargle periodogram \citep[]{scargle82} and the period-searching algorithm of \citet[]{plavchan07}.  The phased light curve is shown in Figure 1a.  For JD=2450000.0, the corresponding phase in Figure 1a is 0.33.  We find a peak-to-peak variation of $\sim$0.4 mag in J, H and K$_s$, with a corresponding reddening of $\sim$0.06 mag in J-H and H-K$_s$ when WL 4 is in the bright state.  While a $\sim$65 day period is consistent with the 2MASS results, our best physical model for the system (${\S}$3.3) requires that this is an alias of the true period.  The period is too long to be associated with a $\sim$1 Myr YSO stellar rotation period \citep[]{rebull01}.  

Super-imposed upon the long-term variability, we identify a second significant period of variability at 4.839$\pm$0.015 days (Figure 1b).  We observe this additional periodic variability in both the faint and bright states, with a larger amplitude in J relative to K$_s$.  This variability is consistent with rotationally modulated stellar variability \citep[]{rebull01}.

\subsection{Spectral Energy Distribution}

The c2d IRAC observations of WL 4 were fortunate to coincide with the bright and faint states predicted from the Cal-PSWDB data.  We model the spectral energy distribution (SED) in both states.  We fit by inspection the photometry with reddened PHOENIX NextGen \citep[]{hauschildt99} synthetic stellar spectra, and a two-temperature blackbody dust model.   For both states of WL 4, we are able to reproduce the observations (Figure 2).  We are able to confirm the presence of an infrared excess associated with a primordial disk.   We summarize the model parameters in Table 1.     

\subsection{Model}

The long period of the near-infrared variability necessitates a binary companion, and the SED indicates the presence a primordial disk.  We denote the binary components WL 4a and WL4b, and the companion resolved in \citet[]{ratzka05} as WL 4c.  All three components are unresolved with 2MASS and \textit{Spitzer}.    The observed IRAC variability implies a circum-binary disk around WL 4ab, but part of the infrared excess could be associated with a disk around WL 4c.  To explain the shape of the light curve, we postulate that a component of the WL 4ab binary goes into obscuration and re-emerges from behind a circum-binary disk every 65.44 days.  The binary must be inclined with respect to the disk, and the disk relatively close to edge-on with respect to our line of sight.  

We can solve for the brightnesses of the three components using the total stellar luminosity derived from the SED fit, the magnitude depth of the faint state from one component being obscured, and the flux ratio observed during a predicted bright state in \citet[]{ratzka05}.  We find that WL 4 is comprised of three approximately equal brightness 0.6 L$_\odot$ YSOs.  The symmetry between the brightnesses of WL 4a and WL 4b implies that the binary period is 130.87 and not 65.44 days.  Our model predicts that WL 4a and WL 4b alternate being obscured by the circum-binary disk.  The similar depth and system color during consecutive faint states is consistent with similar spectral types for WL 4a and WL 4b.   The color variations occur during the ingress and egress phases of the eclipses, and we do not model this additional complexity.  The color changes may indicate a slight difference in spectral types for WL 4a and b, or may be due to varying scattered light flux.  The estimated stellar masses imply a binary separation of 0.47 AU, or $\sim$50 stellar radii.   Finally, we attribute the 4.84 day periodic variability to stellar rotation modulated star-spots on WL 4c, because this is consistent with the variability being observed in both states. 
   
\section{DISCUSSION}

We list supporting evidence for our model in ${\S}$4.1--3.

\subsection{Derived Model Parameters and Lack of Detected Accretion Signatures}

The dust luminosities we derive are sufficient to justify the shadowing model, and the variability is detected from 1--8 $\mu$m.  Only two parameters of our model are necessarily varied between the bright and faint states to reproduce the SED -- the hot dust luminosity which is important to reproduce the IRAC channels 2--4 photometry, and the obscuration of one stellar component in the faint state.  The change in the hot dust luminosity indicates possible dynamical interaction or disk ``warping,'' or simply changing illumination/heating.  The hot dust temperature is consistent with the stable inner orbital radius of $\sim$2--3 times the binary semi-major axis \citep[]{harrington77}.  The presence of WL 4c could account for the dynamical origin of a disk inclined with respect to the orbit of the inner WL 4ab binary.

The lack of a significant component of hot $\sim$1000 K dust and the lack of strong accretion signatures in \citet[]{natta06} imply the lack of a massive circum-primary disk around WL 4a or b.  The lack of detected accretion also implies that periodically driven accretion by the companion is not a favored scenario, such as is hypothesized for DQ Tau and AA Tau \citep[]{matthieu97,bouvier03}.  Finally, the apparent K$_s$ magnitude of WL 4 is consistent with a directly visible $\rho$ Oph YSO, and is too bright to be an edge-on disk system seen only in scattered light \citep[]{stapelfeldt07,watson07,stapelfeldt97}.  

\subsection{Duration of Faint State}

A stable circum-stellar disk around WL 4a or WL 4b with an outer radius of one-third the binary separation \citep[]{artymowicz94} would produce an eclipse only $\sim$10\% of the period, so this configuration can be ruled out.  Excepting for the $\sim$4.84 day periodic variability, the light curves in the faint and bright states are relatively flat and smoothly varying for $\sim$25 and $\sim$13 days respectively. The transition between bright and faint states, including the ``kinks'' in the light curve at phases of 0.14, 0.36, 0.63 and 0.86, last $\sim$13 days apiece.   We speculate that the ``kinks''  could be due to disk substructure such as a puffed up or over-dense inner edge.

\subsection{Similarities to KH-15D}

KH-15D is a unique K6-7 binary YSO in NGC 2264 \citep[]{kearns98,badalian70}.  Every $\sim$48 days, KH-15D periodically varies in brightness by $\sim$4 magnitudes between relatively quiescent bright and faint states.  An inclined binary with a $\sim$48 day period surrounded by a nearly edge-on circum-binary disk is invoked to explain these observations, as the primary component of KH-15D periodically ``peeks out'' from behind the disk \citep[][]{winn06,johnson05,winn04,chiang04,johnson04,winn03}.  The near-IR photometry and colors vary in tandem in a similar fashion for both KH-15D and WL 4, although the amplitude differs.  When both sources get brighter, both sources also get redder in J-H and H-K$_s$ \citep[]{kusakabe05}.   \citet[]{kusakabe05} attribute the reddening for KH-15D to changes in the scattered light flux.  Similar features such as the ``kinks'' in the light curve in Figure 1a have also evolved with time for KH-15D.  

\section{CONCLUSIONS AND FUTURE WORK}

We have identified periodic variability for the YSO WL 4 in $\rho$ Oph that is likely due alternating eclipses of two different components of a binary system by a circum-binary disk.  The faint state lasts longer than the bright state, ruling out a co-planar stable circum-primary disk to explain the observations.  WL 4 is a unique and valuable probe of YSO terrestrial zone disk evolution. We want to confirm our binary model for the light curve modulation.  The radial velocity amplitude implied by our model should be measurable through high-resolution near-IR spectroscopy with adaptive optics.  Detailed modeling of the disk with orbital dynamics of the binary is warranted to reproduce the observed light curves, including the ``kinks'', and to investigate the dynamical stability of our model.  Near- and mid-infrared photometric and spectroscopic observations will enable a characterization of the dust grain properties and disk structure.

\acknowledgements
This publication makes extensive use of data products from 2MASS, which is a joint project of the University of Massachusetts and IPAC/Caltech, funded by NASA and NSF.   This research has made use of the NASA/ IPAC Infrared Science Archive, which is operated by JPL, Caltech, under contract with NASA. Thanks to Mike Meyer, Mike Werner, Angelle Tanner, and Eric Agol for their conversations and comments.   Parts of the research described in this publication was carried out at JPL.

\clearpage

\begin{deluxetable}{ll}
\tablecolumns{2}
\tablewidth{0pc}
\tablecaption{SED Model Parameters}
\tabletypesize{\scriptsize}
\tablehead{
\colhead{Parameter}	& \colhead{Value}
	}
\startdata
\multicolumn{2}{c}{Fixed} \\
\hline
Distance & 135pc  \\
Short $\lambda$ extinction power law & -1.7\tablenotemark{a} \\
Long $\lambda$ extinction power law & -1\tablenotemark{a} \\
Extinction power law transition  & 3.5 $\mu$m\tablenotemark{a}\\
T$_b$,T$_c$,  R$_b$,R$_c$, M$_b$, M$_c$ & T$_a$, R$_a$, M$_a$\tablenotemark{b}\\
\hline
\multicolumn{2}{c}{Varying  -- Best Fit} \\
\hline
WL 4 A$_J$ extinction & 6 mag\\
T$_a$, R$_a$, M$_a$  & 3600 K\tablenotemark{c}, 2.0 R$_\odot$, 0.40 M$_\odot$\tablenotemark{d}  \\
cold dust T, L & 160 K\tablenotemark{e}, 0.017 L$_\odot$ \\
hot dust T, L bright state &  560 K\tablenotemark{e}, 0.055 L$_\odot$ \\
hot dust T, L faint state &  560 K\tablenotemark{e}, 0.061 L$_\odot$\\
\enddata
 \tablenotetext{a}{Extinction wavelength dependence adopted from \citet[]{mathis90,becklin78}.}
 \tablenotetext{b}{An equal-mass triple system model is used (${\S}$3.3).}
 \tablenotetext{c}{Fit to nearest 100 K.}
\tablenotetext{d}{Estimated from 1 Myr 3600K Seiss isochrone. \citep[]{seiss00}.}
\tablenotetext{e}{Fit to nearest 5 K.}
 \end{deluxetable}

\clearpage

\begin{figure}
\plottwo{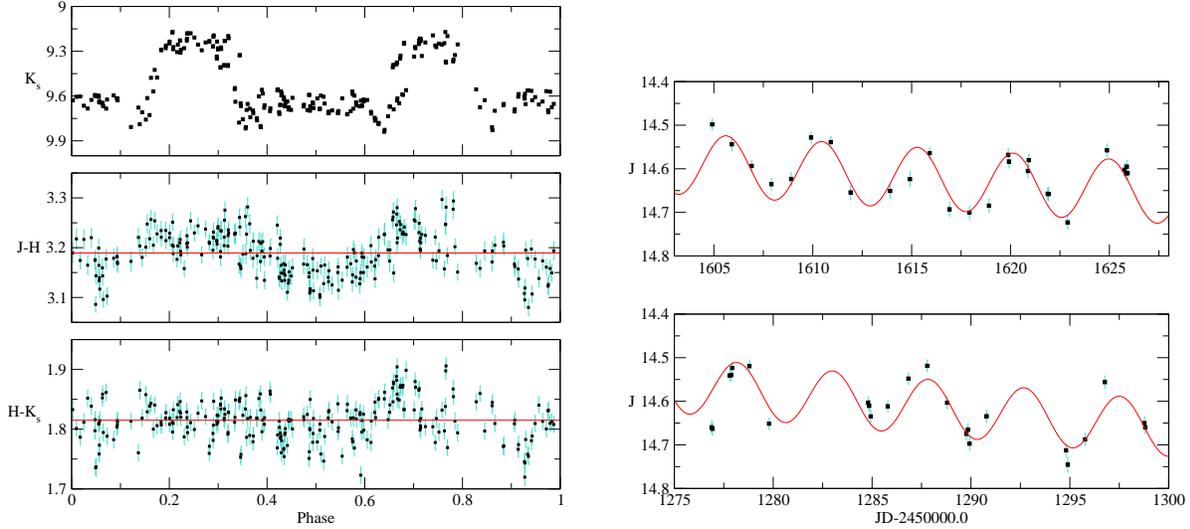}{f1b}
\caption {(a) Left: Top Panel - K$_s$ band Cal-PSWDB light curve data in black for WL 4.  Middle Panel:  J-H Cal-PSWDB color curve.  Bottom Panel:  H-K$_s$ Cal-PSWDB color curve.    Data are folded to a period of 130.87 days, and plotted as a function of period phase.  (b) Right:  Portions of the J-band light curve data during the faint state to show the star-spot variability modulated by the $\sim$5 day stellar rotation period.   Overlaid in red are the best fitting sinusoids plus linear ramps, which reduce the photometric scatter of $\sim$0.07 mag by a factor of $\sim$2.  The residuals are structured and indicate the linear ramps are an over-simplification; we do not identify the physical origin of these variations.  The sinusoids have different phases for all of the spans of faint or bright states that we investigate, implying star-spot evolution with a coherence time-scale longer than the rotation period and on the order of the binary period.  Sinusoidal variations are largest in amplitude at J-band relative to K$_s$-band.  For all panels, each group of six scans from a single hourly calibration observation are co-added, and 1-$\sigma$ error bars are shown in teal.}
\end{figure}

\begin{figure}
\includegraphics[scale=0.3]{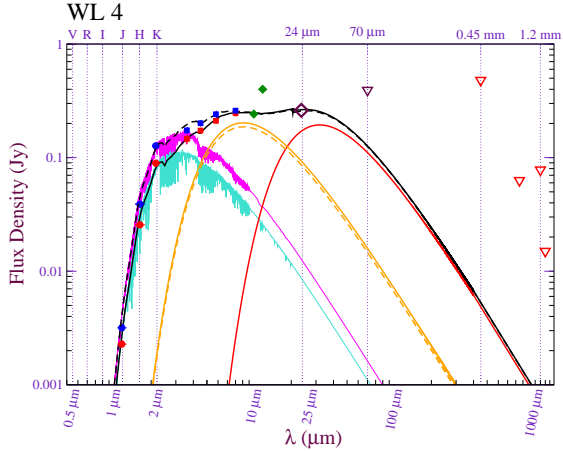}
\caption {Spectral Energy Distribution for WL 4.  DATA: Red circles - 2MASS Cal-PSWDB flux densities in the faint state (Figure 1a phase 0); blue circles - 2MASS Cal-PSWDB flux densities in the bright state (phase 0.25); red squares - c2d IRAC flux densities in the faint state (JD=2453071.831, Figure 1a phase 0.142); blue squares - c2d IRAC flux densities in the bright state (2453092.736, 0.302); green diamonds - 10.8 and 12.5$\mu$m measurements \citep[$\sim$2450198.0, 0.183; $\sim$2451358.0, 0.047,][]{barsony05}; purple diamond - c2d MIPS 24 $\mu$m flux density at both epochs (2453083.888, 0.235; 2453084.113, 0.236); purple triangle - c2d MIPS 70 $\mu$m flux density upper limit; red triangles - sub-mm flux density upper limits \citep[]{andrews07,stanke06}.   The IRAC epochs correspond to the channel 1 and 3 observations.  The channel 2 and 4 observations were taken within one hour of these times.  MODEL: solid red line - cold dust component; solid orange line -  hot dust component in the faint state; dashed orange line - hot dust component in the bright state; cyan line - reddened synthetic spectra for the two WL 4 components un-obscured in the faint state; magenta line - reddened synthetic spectra for the three WL 4 components un-obscured in the bright state; black solid line - sum of stellar components, hot dust and cold dust spectra in faint state; black dashed line - sum of stellar components, hot dust and cold dust spectra in bright state.}
\end{figure}

\end{document}